\documentclass[prl,nofootinbib,showpacs,superscriptaddress,preprintnumbers,amsmath,amssymb]{revtex4}

\usepackage{graphicx, color}
\usepackage{dcolumn}
\usepackage{bm}
\usepackage{amsmath}

\renewcommand\Re{\mathop{\mathrm{Re}}}

\newcommand\p{\partial}

\begin{document}

\title{
Dynamics of vortices in chiral media: the chiral propulsion effect
}

\author{Yuji~Hirono}
\email{yhirono@bnl.gov}
\affiliation{Department of Physics, Brookhaven National Laboratory,
Upton, New York 11973-5000, USA}
\author{Dmitri~E.~Kharzeev}
\affiliation{Department of Physics and Astronomy, Stony Brook University, Stony Brook,
 New York 11794-3800, USA}
 \affiliation{Department of Physics and RIKEN-BNL Research Center, Brookhaven National Laboratory,
Upton, New York 11973-5000, USA}

\author{Andrey~V.~Sadofyev}
\affiliation{
Theoretical Division, MS B283, Los Alamos National Laboratory, Los Alamos, NM 87545
}

\date{\today}

\begin{abstract}
We study the motion of vortex filaments in chiral media, and find a
semi-classical analog of the anomaly-induced chiral magnetic
effect.
The helical solitonic excitations on vortices in a
parity-breaking medium are found to carry an additional energy flow
along the vortex in the direction dictated by the sign of chirality
imbalance; we call this new transport phenomenon the Chiral Propulsion
Effect (CPE).
The dynamics of
the filament is described by a modified version of the localized induction
equation in the parity-breaking background. 
We analyze the linear stability of simple vortex
configurations, and study the effects of chiral media on the
excitation spectrum and the growth rate of the unstable modes.
It is also shown that, if the equation of motion of the filament is symmetric 
under the simultaneous reversal of parity and time, the resulting planar
solution cannot transport energy.
\end{abstract}

\maketitle


\section{Introduction}

The physics of chiral media has attracted a significant attention
recently. 
Remarkably, it appears that the quantum chiral anomaly~\cite{adler1969axial,bell1969pcac} significantly 
affects the macroscopic behavior of chiral media and induces new transport phenomena, such as 
the Chiral Magnetic
\cite{Kharzeev:2004ey, Kharzeev:2007jp,Fukushima:2008xe,Kharzeev:2015znc,Kharzeev:2013jha}
and Chiral Vortical Effects
\cite{Kharzeev:2007tn, Erdmenger:2008rm, Banerjee:2008th, Torabian:2009qk, Son:2009tf}
(CME and CVE, respectively). CME and CVE refer to the generation of 
electric currents along an external magnetic field or vorticity 
in the presence of a chirality imbalance.
The resulting currents are non-dissipative due to
the protection  by the global topology of the gauge field.
These chiral effects are expected to occur in a variety of systems: the quark-gluon
plasma,  Dirac and Weyl semimetals,  primordial electroweak plasma,  and
cold atoms. In  quark-gluon plasma, the chirality imbalance can be
produced by topological fluctuations of QCD, or by the combination of
electric and magnetic fields that accompany heavy-ion collisions. The
parallel electric and magnetic fields can also be used to create the
chirality imbalance in condensed matter systems, see
e.g. \cite{Li:2014bha}.
In addition, 
CME and CVE lead to a new class of instabilities in these systems
\cite{Joyce:1997uy, Akamatsu:2013pjd, Avdoshkin:2014gpa,Tashiro:2012mf,
Hirono:2015rla, Yamamoto:2016xtu, Hattori:2017usa}.

The CME has been observed experimentally in Dirac \cite{Li:2014bha,
xiong2015evidence, li2015giant} and Weyl semimetals
\cite{zhang2015observation, yang2015chiral, yang2015observation}.
There is an ongoing search for CME and the local parity violation
\cite{Kharzeev:2004ey, Kharzeev:2007jp} induced by the topological
fluctuations in the quark-gluon plasma in heavy-ion collisions at RHIC
and LHC, see Ref.~\cite{Kharzeev:2015znc} for a review. In particular, the
forthcoming isobar run in the Spring of 2018 at RHIC is expected to
provide a conclusive result on the occurrence of CME in heavy-ion
collisions \cite{Skokov:2016yrj}.

Recently, the STAR collaboration reported the experimental observation
of $\Lambda$ hyperon polarization along the normal to the reaction plane
of the heavy-ion collision, pointing towards the existence of large
vorticity in the produced quark-gluon fluid \cite{STAR:2017ckg}.
The role of vortical flows in heavy-ion collisions has been 
discussed e.g. in Refs.~\cite{Liang:2004ph,Betz:2007kg,Becattini:2013vja,Becattini:2015ska, Baznat:2015eca, Pang:2016igs,Deng:2016gyh,Jiang:2016woz,Becattini:2016gvu,Karpenko:2016jyx,Li:2017dan,Aristova:2016wxe}. 
It is natural to ask how the dynamics of vortices is influenced by
the chiral anomaly. 

In this paper, we consider the dynamics of vortices in a fluid with
broken parity; the electromagnetic fields are treated as fully
dynamical.
We find a new chiral transport effect  -- an additional, asymmetric
energy flow along the vortex filament in the direction determined by the
sign of chirality imbalance, the Chiral Propulsion Effect (CPE).

\section{Excitations on Vortices in chirally imbalanced media}

Consider the motion of a vortex filament in a fluid. It can be described by the localized
induction equation (LIE)\footnote{Although originally the LIE has been 
introduced for thin vortices in classical fluids
\cite{saffman1992vortex, ricca1991rediscovery}, it can also describe the
dynamics of quantized vortices in superfluids and superconductors
\cite{vilenkin2000cosmic, volovik2009universe, Eto:2013hoa}.},
\begin{equation}
\dot{ \bm X } = C \bm X' \times \bm X'' , 
 \label{eq:lie}
\end{equation}
where $\bm X = \bm X(t,s)$ denotes the position of a vortex, $t$ is the
time, $s$ is the arc-length parameter,
the dot and the prime indicate the 
derivatives with respect to $t$ and $s$ respectively,
and $C$ is a parameter dependent on the properties of the fluid. 
Interestingly, 
the LIE (\ref{eq:lie}) can be mapped to the non-linear
Schr\"{o}dinger equation (NLSE) by the so-called Hasimoto transformation
\cite{hasimoto1972soliton},
\begin{equation}
 \psi(t,s) = \kappa(t,s) \exp \left[i \int^s \tau(t,s') ds'  \right],
 \label{eq:hasimoto-tr}
\end{equation}
where $\kappa(t,s)$ is the curvature and $\tau(t,s)$ is the torsion of a
vortex.
NLSE is known to be a completely integrable
system which has solitonic solutions and an infinite sequence of
commuting conserved charges. The Hasimoto transformation has been shown to be
a Poisson map that preserves the Poisson structures
\cite{langer1991poisson}. The LIE thus describes a completely integrable
system. The LIE possesses solutions that represent helical excitations propagating along the vortex; they are known as Hasimoto solitons.

Let us now consider a system in which parity is broken by the presence of magnetic helicity; the corresponding term in the action is
\begin{equation}
S_{\chi} = \int dt  \,  \mu \, \mathcal{H},
  \label{eq:s-cs}
\end{equation}
where
$\mu$ is the ``chiral'' chemical potential\footnote{Usually it is denoted by $\mu_5$, but as this will be the only chemical potential that will appear in this paper, we simplify this notation to $\mu$. Please note that the chiral chemical potential does not correspond to a conserved quantity, as the chiral charge is not conserved due to the chiral anomaly. The state with $\mu \neq 0$ therefore does not correspond to a true ground state of the system; for discussion, see e.g. \cite{Kharzeev:2013ffa}.}, 
$\mathcal{H}$ is the magnetic helicity given by
$\mathcal H = \frac{e^2}{4 \pi^2} \int d^3 x \, \bm A \cdot \bm B$,
where $\bm A$ is the
vector potential and $\bm B$ is the magnetic field.
It is worth mentioning 
that taking the derivative of this action with respect to the
vector potential, one readily finds the CME current: $\bm J_{\rm CME} = \delta
S_\chi / \delta \bm A \propto \bm B$. Supplementing the non-relativistic
Abelian Higgs
model with the term given by Eq.~(\ref{eq:s-cs}),  one can find the equation of motion
for a quantized magnetic vortex at finite $\mu$, as derived by Kozhevnikov 
\cite{Kozhevnikov:1999ak,Kozhevnikov:2015oga} :
\begin{equation}
 \dot{\bm X}
  = 
C \bm X ' \times \bm X''
+ \mu
\left[ \bm X'''
 +
 \frac{3}2
 (\bm X'')^2 \bm X'  \right] , 
\label{eq:fm-eq}
\end{equation}
where a tangential term $ \frac{3}{2} \mu (\bm X'')^2 \bm X' $ is added
to keep the arc-length-preserving property\footnote{
The term $\bm X'''$ can also be derived in a fluid-dynamical system
using the kinetic helicity as the Hamiltonian~\cite{holm2004hasimoto}. 
}.
Note that the tangential motion does not change the shape of the vortex.
Hereafter, we set the constant $C$ in Eq.~(\ref{eq:fm-eq}) to
unity by a corresponding time rescaling. 
Let us note that Eq.(\ref{eq:fm-eq}) has previously emerged in a different context: 
it describes the motion of a vortex tube containing an axial flow, and is known as
 the Fukumoto-Miyazaki
equation (FME) \cite{fukumoto1991three, kambe2004geometrical}. 
Remarkably, through the Hasimoto transformation,
the FME
can be mapped to the integrable  Hirota equation~\cite{hirota1973exact}, 
\begin{equation}
 i \dot \psi + \psi'' + \frac{1}2 |\psi|^2 \psi
  - i  \mu \left(
\psi''' + \frac{3}{2} |\psi|^2 \psi' 
		      \right) = 0 ;
 \label{eq:hirota-eq}
\end{equation}
this map can be utilized to obtain the solitons of the FME. 

In this paper we are interested in the behavior of chiral solutions. 
We can find a simple explicit solution of the FME~(\ref{eq:fm-eq})
having the form of a helix, 
\begin{equation}
 {\bm X}_{\rm helix}(t,s) =
  \frac{1}{A^2}
  \begin{pmatrix}
   \kappa_0 \cos[ A (s - v_p t) ] \\
   \kappa_0 \sin[ A (s - v_p t) ] \\
    \tau_0 \, A (s - v_g t) \\
  \end{pmatrix}, 
\label{eq:sol-helix}
\end{equation}
where the constants $\kappa_0$ and $\tau_0$ give the curvature and the torsion of the helix, 
$
 A = \sqrt{\kappa_0^2 + \tau_0^2 } , 
$
and the phase and group velocities are given by 
$
 v_p = \tau_0 + \mu (\tau_0^2 - \frac{\kappa_0^2}{2}), 
$
$
 v_g = - \frac{\kappa_0^2}{\tau_0} - \frac{3 \kappa_0^2}{2} \mu . 
$
Note that the sign of $\tau_0$ determines the handedness of the helix. The radius $R$ and the pitch $\ell$ of the helix are given by 
$
 R = \kappa_0 / ( \kappa_0^2 + \tau_0^2 ), \ 
\ell = 2 \pi \tau_0 /(\kappa_0^2 + \tau_0^2 )
$. 
The solution is reduced to a circular loop in the limit $\tau_0=0$
in Eq.~(\ref{eq:sol-helix}).

Using the map between the FME and the Hirota equation, 
we find a propagating solitonic solution of the FME, 
\begin{equation}
 {\bm X}_{\rm sol}(t,s) = 
  \begin{pmatrix}
   - \frac{2 \epsilon}{\epsilon^2 + \tau_0^2} {\rm \ sech  } [\epsilon \xi]
   \cos \left[ \eta \right] \\
   - \frac{2 \epsilon}{\epsilon^2 + \tau_0^2} {\rm \ sech  } [\epsilon \xi]
   \sin\left[ \eta \right] \\
   s  - \frac{2 \epsilon}{\epsilon^2 + \tau_0^2} \tanh [\epsilon
   \xi]
  \end{pmatrix} .
  \label{eq:sol-soliton}
\end{equation}
where
$
\eta \equiv \tau_0 s + (\epsilon^2 - \tau_0^2)t + \mu \tau_0(3 \epsilon^2 -
 \tau_0^2),
$
$
\xi \equiv s - (2 \tau_0 + \mu \left( 3 \tau_0^2 - \epsilon^2 \right) )t,
$
and $\epsilon$ and $\tau_0$ are constants. This soliton has a constant
torsion given by $\tau_0$ and propagates in the $z$ direction. Its
speed is modified by $\mu$ and reduces to the original Hasimoto soliton
at $\mu = 0$.

Let us discuss the kinetic properties of these solutions. The kinetic
energy of a soliton can be found as
$
  E = \frac{1}2 \int ds \, \kappa^2  = 4 \epsilon . 
$
In addition, the helical nature of the configuration \cite{ricca1992physical} can be characterized by the quantity 
$
 {\mathcal H} = \int ds \, \kappa^2 \tau = 8 \epsilon \tau_0 .
$ 
This is the second conserved quantity in the NLS hierarchy
\cite{langer1991poisson, ricca1992physical}.  In the case of a planar
configuration, namely if the torsion is vanishing, ${\mathcal H}=0$. Note that
these quantities do not depend on $\mu$.

Let us now turn to the momentum carried by these solutions. 
In the thin vortex limit, the electromagnetic fields can be expressed in terms of the vortex coordinates $\bm X(t,s)$ as 
\begin{equation}
 \bm B(t, \bm x) =  \varphi \int d s \, \bm X' \delta(\bm x - \bm X), 
\end{equation}
\begin{equation}
 \bm E(t, \bm x) =
  -  \varphi \int d s \, \dot{\bm X} \times \bm X' \delta(\bm x - \bm X) , 
\end{equation}
where $\varphi$ is the magnetic flux and the electric field locally has the structure ``$\bm v \times \bm B$.''
The momentum of the magnetic flux is given by the Poynting vector, 
\begin{equation}
 \bm P = \int d^3x \, \bm E \times \bm B
  = \varphi^2 M^2 \int ds \, \dot{\bm X} ,
\end{equation}
where $M$ is the inverse of the core size of the vortex.
The helix solution moves in the $z$-direction. 
The $z$-component of momentum per unit length of the coil is evaluated
as
\begin{equation}
 ( {\bar{P}}_{\rm helix}
  )_z = \frac{\kappa_0^2}{\tau_0}\left(1 + \frac{3}{2} \mu \tau_0
				       \right) \varphi^2 M^2 .
  \label{eq:pz-helix}
\end{equation}
The $z$ component of the momentum of the soliton solution can be calculated
using Eq.~(\ref{eq:sol-soliton}):
\begin{equation}
 (P_{\rm sol})_z = \frac{4 \epsilon(2 \tau_0 + \mu(3 \tau_0^2 - \epsilon^2) )
  }{\epsilon^2 + \tau_0^2} \, \varphi^2 M^2 .
  \label{eq:pz-sol}
\end{equation}
In both cases, there are contributions proportional to
$\mu$. Therefore the chiral medium provides a thrust to the solitons, propelling them along the vortex - we will call this  
 the Chiral Propulsion Effect (CPE).
 
In the case of the solitons (\ref{eq:pz-sol}), at $\mu=0$ 
the velocity is proportional to the torsion $\tau_0$ -- this means that for the wave to have a finite momentum in a chirally symmetric medium, 
the vortex has to deform in a parity-breaking way. On the other hand,
even if the solution is planar, due to the circulation in the vortex it
can still experience the thrust if parity is broken in the
medium. Indeed, Eq (\ref{eq:pz-sol}) shows that for $\mu \neq 0$ the
thrust remains even in the $\tau_0 \to 0$ limit corresponding to a
planar solution with ${\mathcal H} = 0$.
As we will discuss later, a planar solution is forbidden to have a finite energy
flow in a PT symmetric theory. 
The LIE has the PT symmetry, while in the FME case it is broken.

\section{Properties of fluctuations}

Let us now examine the effect of the chiral medium on the fluctuations around the
circle and helix solutions.
We use the local coordinate system called the Frenet--Serret (FS) frame,
which is commonly used to parametrize the shape of a curve. 
There is an ambiguity in the parametrization in $s$, and we fix this by
requiring $|\bm X' |= 1$.  Then, the unit tangent vector is written as
$\bm t = \bm X'$.
Given a curvature $\kappa(t,s)$ and a torsion $\tau(t,s)$, the shape of
a curve is determined,  up to a trivial translation and rotation, by the
FS formulas,
\begin{equation}
 \p_s
\begin{pmatrix}
 \bm t \\
 \bm n \\
 \bm b \\
\end{pmatrix}
=
\begin{pmatrix}
0 & \kappa & 0 \\
-\kappa  & 0 & \tau \\
0 & -\tau & 0 \\
\end{pmatrix}
\begin{pmatrix}
 \bm t \\
 \bm n \\
 \bm b \\
\end{pmatrix} ,
\end{equation}
where $\bm n \propto \bm t'$ is the unit normal vector, and $\bm b \equiv \bm t \times \bm n$ is the unit binomal vector. The time evolution of a curve is described by 
\begin{equation}
\p_t
 \begin{pmatrix}
\bm t \\
\bm n \\
\bm b \\
 \end{pmatrix}
 =
 \begin{pmatrix}
  0   & \alpha & \beta \\
-\alpha   & 0 & \gamma \\
- \beta  & -\gamma & 0 \\
 \end{pmatrix}
  \begin{pmatrix}
\bm t \\
\bm n \\
\bm b \\
  \end{pmatrix} , 
  \label{eq:eom-fm}
\end{equation}
where $\alpha, \beta, \gamma$ are functions of $\kappa$ and $\tau$ and
their functional forms are determined from Eq.~(\ref{eq:fm-eq}). 
The FS basis has to satisfy the compatibility conditions,
$
 \p_s \p_t \bm t= \p_t \p_s \bm t, \
 \p_s \p_t \bm n= \p_t \p_s \bm n, \
 \p_s \p_t \bm b= \p_t \p_s \bm b
$.
Using these conditions, we find (see the Supplementary Material)
the time-evolution equations for $\tau$ and $\kappa$:
\begin{equation}
 \dot \kappa =
  -2 \tau  \kappa ' -\kappa  \tau '  
+
  \frac{\mu }{2}  \left(2 \kappa''' -6 \tau ^2 \kappa '+3 \kappa ^2 \kappa '-6 \kappa  \tau 
   \tau '\right) , 
  \label{eq:darios-fme-kappa}
\end{equation}
\begin{equation}
\begin{split}
 \dot \tau &=
\frac{\kappa''' \kappa -2 \kappa ^2 \tau  \tau '+\kappa ^3 \kappa
 '-\kappa ' \kappa ''}{\kappa ^2} \\
 & +
 \frac{\mu} {2 \kappa ^2} 
   \left(3 \kappa ^4 \tau '+6 \kappa''' \kappa  \tau +2 \kappa ^2
 \tau''' +12 \kappa  \kappa '' \tau '-6
   \kappa ^2 \tau ^2 \tau '+6 \kappa  \kappa ' \tau ''-6 \left(\kappa '\right)^2 \tau '+6 \kappa ^3 \tau  \kappa
 '-6 \tau  \kappa ' \kappa ''\right)
  . 
\end{split}
\label{eq:darios-fme-tau}
\end{equation}
If we take $\mu=0$ in Eqs.~(\ref{eq:darios-fme-kappa}) and (\ref{eq:darios-fme-tau}) the Da Rios equations are reproduced
\cite{da1906motion}.

We consider linear fluctuations, $\delta \kappa$ and  $\delta \tau $, 
around constant $\kappa$ and $\tau$.
By taking $\delta \kappa,\ \delta \tau  \propto e^{-i \omega t + i ps }$, 
the following dispersion relation is obtained from
Eqs.~(\ref{eq:darios-fme-kappa}) and (\ref{eq:darios-fme-tau}), 
\begin{equation}
\omega = 
 2 p \tau
 + \mu p 
 \left(
 p^2
 -\frac{3}2 \kappa ^2   +3  \tau ^2
 \right)
 \pm 
 \sqrt{
 p^2( p^2- \kappa^2)
(1 + 3 \mu \tau )^2
 } . 
\label{eq:omega}
\end{equation}
Equation (\ref{eq:omega}) compactly encodes the information of the
fluctuations around three different configurations: a circle, a helix
and a straight line.
Let us first discuss a circle, in which case the torsion is zero. 
The periodicity of a circle requires $p = n \kappa$ with an integer
$n$, then the frequency $\omega$ simplifies to 
\begin{equation}
 \omega =\pm \kappa^2 \sqrt{ n^2( n^2- 1)}
  + \mu \kappa^3 \, n \left(n^2 - \frac{3}{2} \right). 
\label{eq:omega-circle}
\end{equation}
If we take $\mu =0$, Eq.~(\ref{eq:omega-circle}) coincides with the
result of previous studies
\cite{kambe1971motion, betchov1965curvature,
ricca1996contributions,PhysRevLett.77.1679}.
The mode with $n=0$ corresponds to the change of radius, 
$n=\pm 1$ represents a slight change of the propagation
direction, and does not involve the change of its shape.
At $\mu=0$, these modes are the zero modes of the soliton.
At $\mu \neq 0$, because of the chirality imbalance,
$n=\pm 1$ modes acquire finite frequencies.
The degeneracy between $n \leftrightarrow -n $ is also lifted. 
The frequency is always real, which means that a circle is 
stable. 

Let us now consider the helix-shape solution. 
The lowest value of $p$ is determined by the length $L$ of a helix as $2
\pi /L$. 
At $\mu=0$, the imaginary part appears if $p^2 < \kappa^2$, which
means that these long-wavelength modes are unstable. 
Since the factor $(1 + 3 \mu \tau )^2$ is always nonnegative, 
this condition is unchanged, except
for a very special choice of the chiral chemical potential $\mu \tau= -1/3$. 
However, a finite $\mu$ changes the growth rate of unstable modes.
In the small $p$ limit, the growth rate is given by 
\begin{equation}
 {\rm Im}\, \omega = \pm \sqrt{\kappa^2 (1+3 \mu \tau)^2 }\, p  +
  O(p^2), 
\end{equation}
which is different for the right-handed ($\tau>0$) and
left-handed ($\tau<0$) helices. 
The real part of $\omega$ to the first order in $p$ is given by
\begin{equation}
{\Re \omega} = \left[ 2 \tau + \frac{3}2 \mu (2 \tau^2 - \kappa^2)
		 \right] p  + O(p^2)  . 
\end{equation}
Hence, the chirality imbalance also modifies the velocity of the wave
propagating along the helix. 

In the limit $\kappa \rightarrow 0, \tau \rightarrow 0$, the helix
approaches a straight line \cite{ricca1994effect}, 
and 
the dispersion for the fluctuations around a straight vortex, $\omega
=\pm p^2 + \mu p^3$, is obtained. 
The leading $p^2$ behavior corresponds to the famous Kelvin waves 
\cite{thomson1880xxiv}, and the second term represents the modification
due to a chirality imbalance. 

\begin{figure}[tbp]
\centering
\includegraphics[width=0.5\textwidth]{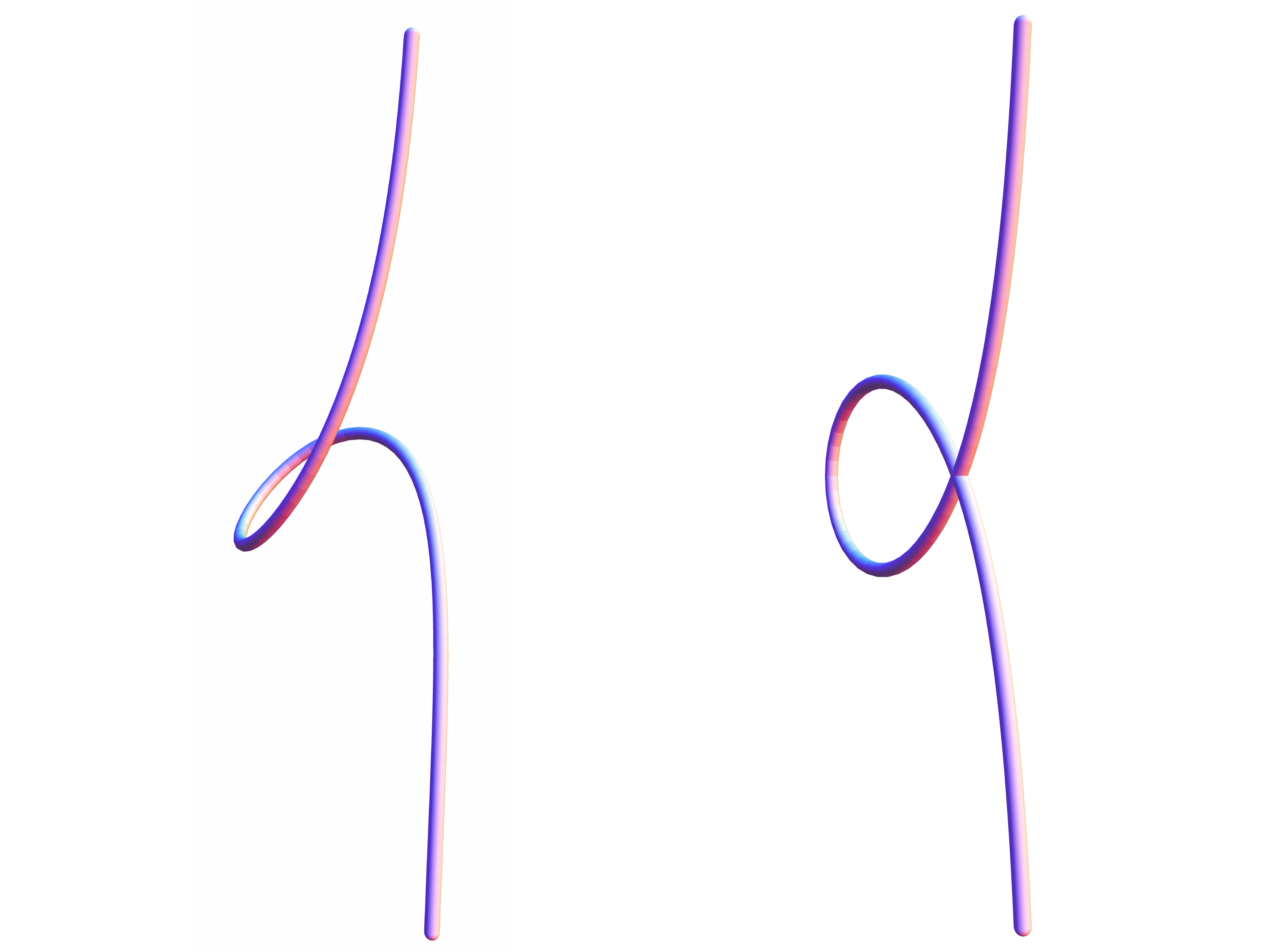}
\caption{
Non-planar (left, $(\epsilon,\tau_0,t)=(1,0.4,0)$) and planar (right,
$(\epsilon,\tau_0,t)=(1,0,0)$) solitons. 
}
\label{fig:sol}
\end{figure}

\section{ Absence of propagation of planar solutions in the PT-symmetric case}

In the case of the LIE, the velocity of a Hasimoto soliton is given by
$v = 2\tau_0$ (take $\mu=0$ in Eq.~(\ref{eq:sol-soliton})). 
At $\tau \ne 0$, the solutions are ``chiral,'' in the sense that their
handedness is correlated with the direction of the propagation. 
A mirror image of a solution propagates in the opposite direction from
the original one.
If we look at a planar ($\tau=0$) solution at $\mu=0$ (see the right figure of
Fig.~\ref{fig:sol}), it 
just rotates around its axis and cannot convey energy along the vortex. 
In fact, this is a generic feature. 
Here, we consider a class of solutions that are asymptotically 
straight lines, like the Hasimoto solitons. 
We will now show that these solutions cannot propagate if the equation of motion (EOM) has the
PT symmetry. 
The LIE has this symmetry, while the FME does not. 

It suffices to show that, when the solution is planar, 
the velocity $\dot{\bm X}$ is restricted to the direction of $\bm b$, since
a binomal motion cannot make the soliton propagate along the vortex. 
Consider a current written in the form 
\begin{equation}
 \bm J = \int ds \, f(\kappa, \tau) \dot{\bm X}, 
\end{equation}
where $f(\kappa, \tau)$ is a function of $\kappa$ and $\tau$.
The energy current is written in this form. 
Let us denote the unit vector in the direction of the asymptotic line by
$\bm \ell$.
Since $\bm \ell$ is within the plane spanned by $\{ \bm t, \bm n\}$ for
a planar solution, 
it is always orthogonal to $\bm b$, $\bm \ell \cdot \bm b = 0$.
Thus, if $\dot{\bm X}\propto \bm b$, 
then $\bm \ell \cdot \bm J=0$ holds and there is no energy flow in the
direction of $\bm \ell$. 

Let us examine the transformation property of the EOMs.
The parity reflection, $\bm X \rightarrow - \bm X$, 
acts on the FS system as
\begin{equation}
\{ \bm t, \bm n, \bm b, \kappa, \tau \}
 \longrightarrow
\{ - \bm t, - \bm n, \bm b, \kappa,- \tau \} . 
\end{equation}
The binomal vector is parity even, because $\bm b = \bm t \times \bm n$,
while the torsion $\tau = (\bm n' \cdot \bm b)$ is parity odd.
The RHS of the LIE, $\bm X' \times \bm X''$, is P-even, while
the modification to the LIE in the FME,
$\bm X'''+\frac{3}{2} (\bm X'')^2 \bm X'$, is P-odd. 

A general EOM can be written in the form
\begin{equation}
 \dot{ \bm X}
  = a(\kappa, \tau) \, \bm t
  + b(\kappa, \tau) \, \bm n
  + c(\kappa, \tau) \, \bm b .
\label{eq:abc}  
\end{equation}
From our assumption, the theory has the PT symmetry.
The LHS is even under PT. 
The RHS is T-even, so it has to be P-even. 
For a planar solution, $\tau=0$, the coefficients in
Eq.~(\ref{eq:abc}) are all P-even, because $\kappa$ is a P-even
quantity. 
Thus, the coefficients of $\bm t$ and $\bm n$ have to vanish,
$a(\kappa,\tau=0) = b(\kappa, \tau=0) = 0$, and
$\dot{\bm X} \propto \bm b$. 

The results above can be further generalized. 
The LIE can be mapped to NLSE, which 
has an infinite sequence of commuting invariants. 
Those invariants are the generators of the Hamiltonian flows.
Correspondingly, the LIE also has infinitely many commuting Hamiltonian
flows \cite{langer1991poisson}, which are called the LIE hierarchy. 
The first and second term of the RHS of the FME (\ref{eq:fm-eq}) are 
the first two Hamiltonian flows, 
\begin{equation}
\bm V_0 = \kappa \bm b,
 \quad
 \bm V_1 = \frac{\kappa^2}2 \bm t + \kappa' \bm n + \kappa \tau \bm b ,
\quad 
\cdots
\end{equation}
In Ref.~\cite{langer1991poisson}, a recursion operator that successively
generates the next flow is constructed, 
\begin{equation}
 \mathcal R \bm V \equiv
  - \mathcal P \left[ \bm t \times \p_s \bm V \right], 
\end{equation}
where $\mathcal P$ denotes the reparametrization procedure to keep the
arc-length-preserving nature, which is done by adding a tangential
term (see also Ref.~\cite{holm2004hasimoto}).
Once we know $\bm V_{n}$, we can obtain the next flow by 
$
\bm V_{n+1} =  \mathcal R \bm V_{n}
$. 
One can show (see the Supplementary Material) that 
$\bm V_n$ is P-even(odd) if $n$ is an even(odd) number. 
Thus, every EOM with an even $n$ has the PT symmetry, and the 
solution of the EOM cannot propagate if its planar.

\vskip0.3cm

To summarize, we have found a new phenomenon affecting the dynamics of
vortex solitons in chirally imbalanced media - the Chiral Propulsion
Effect. The CPE refers to
an asymmetric energy flow along the vortex filament in the direction
determined by the sign of the chirality imbalance. The energy is carried
along the vortex by helical excitations analogous to the
Hasimoto solitons. We have also found that the growth rate of unstable
modes on the helical soliton solution is modified by the chirality of
the medium.
It is shown that, if the equation of motion respects the PT
symmetry, a planar
solution cannot transfer energy --  
this indicates that the existence of the CPE is entirely due to the
breaking of parity in the medium.

\vskip0.3cm

\begin{acknowledgements}
This work was supported in part by the
U.S. Department of Energy 
under contracts No. DE-FG-88ER40388 and DE-SC-0017662 (D.K.), DE-AC02-98CH10886 (Y.H. and D.K.),
and within the framework of the Beam Energy Scan Theory (BEST)
Topical Collaboration. The work of A.S. is partially supported through the LANL LDRD Program.
\end{acknowledgements}

\section{Supplementary Material}

\subsection{Derivation of Eqs. (\ref{eq:darios-fme-kappa}) and
(\ref{eq:darios-fme-tau})}

The compatibility conditions for the Frenet-Serret frame, 
$
 \p_s \p_t \bm t= \p_t \p_s \bm t, \
 \p_s \p_t \bm n= \p_t \p_s \bm n, \
 \p_s \p_t \bm b= \p_t \p_s \bm b
$, 
result in the following relations, 
\begin{equation}
 \p_s \beta =- \alpha \tau + \gamma \kappa  , 
\end{equation}
\begin{equation}
 \p_t
  \begin{pmatrix}
   \kappa \\
   \tau
  \end{pmatrix}
  =
    \begin{pmatrix}
   0 & - \beta \\
 \beta & 0 
    \end{pmatrix}
 \begin{pmatrix}
  \kappa \\
  \tau
 \end{pmatrix}
 +
 \p_s
  \begin{pmatrix}
  \alpha \\
  \gamma
  \end{pmatrix} .
  \label{eq:com-2}
\end{equation}
A generic equation of motion (EOM) of a curve is written in the form  
\begin{equation}
 \dot{ \bm X} = a(\kappa,\tau) \, \bm t + b(\kappa,\tau) \, \bm n +
  c(\kappa,\tau) \, \bm b , 
  \label{eq:abc-1}
\end{equation}
where $a(\kappa,\tau),b(\kappa,\tau),c(\kappa,\tau)$ are functions of $\kappa$ and $\tau$. They are related to $\alpha, \beta, \gamma$ in Eq.~(\ref{eq:eom-fm}) as 
\begin{equation}
 \alpha = a \kappa + b' - c \tau,
\quad
 \beta = b \tau + c',
\quad
 \gamma = \frac{\beta' + \tau \alpha} {\kappa}, 
  \label{eq:comp-alpha}
\end{equation}
which can be checked by using the FS formulas.
The FME (\ref{eq:fm-eq}) can be written
in terms of the FS basis as 
\begin{equation}
  \dot{ \bm X}
  =  \frac{\mu}{2} \kappa^2 \bm t + \mu \kappa' \bm n + \kappa(1 +
  \mu \tau) \bm b . 
\end{equation}
From this expression we can read off $a$, $b$ and $c$ in
Eq.~(\ref{eq:abc-1}).
By plugging them into 
Eqs.~(\ref{eq:comp-alpha}),
we obtain the expressions for $\alpha, \beta, \gamma$. Substituting them into Eq.~(\ref{eq:com-2}),
we find the time-evolution equations for $\tau$ and $\kappa$ given by Eq.(\ref{eq:darios-fme-kappa}) and Eq.(\ref{eq:darios-fme-tau}).

\subsection{The parity symmetry of the higher-order flows of the LIE hierarchy}

Here we determine the parity symmetry of the higher-oder flows of the
LIE hierarchy, using the recursion operator $\mathcal R$. 
It is shown that 
$\bm V_n$ is P-even(odd) if $n$ is an even(odd) number. 
Suppose $\bm V_n$ is a flow of a particular parity (even or odd).
The $n+1$-th flow can be
generated by the operation, 
\begin{equation}
\begin{split}
 \bm V_{n+1} &=
 - \mathcal R (\bm X' \times \p_s \bm V_n) \\
 &= - \bm X' \times \p_s \bm V_n + \bar a(\kappa, \tau) \, \bm t ,
\end{split}
\end{equation}
where $\bar a(\kappa, \tau) \, \bm t$  is the added term to keep
the arc-length unchanged. 
The multiplication of $\bm X' \times \p_s $ changes the parity, because
of a factor of $\bm X'$, and the first term has the opposite parity from
$\bm V_{n}$. 
One can see that the reparametrization operation does not change the
parity of the flow, as follows. 
$\bm V_{n+1}$ can be written in the form, 
$
 \bm V_{n+1} =
   b(\kappa,\tau) \, \bm n
  + c(\kappa,\tau) \, \bm b
  + \bar a(\kappa,\tau) \, \bm t . 
$
The arc-length preserving condition, 
$ \bm t \cdot \bm V_{n+1}' = 0$, 
implies that the newly added term has to satisfy 
$
 \bar{a}' =\kappa b. 
$
Since $\kappa$ is P-even,
$\bar{a}$ has the same parity as $b$, and $\bar a \, \bm t$ has the same
parity as $b \, \bm n$.
Therefore, $\bm V_n$ is P-even(odd) if $n$ is an even(odd) number.

\bibliography{refs-1}

\begin{thebibliography}{61}
\expandafter\ifx\csname natexlab\endcsname\relax\def\natexlab#1{#1}\fi
\expandafter\ifx\csname bibnamefont\endcsname\relax
  \def\bibnamefont#1{#1}\fi
\expandafter\ifx\csname bibfnamefont\endcsname\relax
  \def\bibfnamefont#1{#1}\fi
\expandafter\ifx\csname citenamefont\endcsname\relax
  \def\citenamefont#1{#1}\fi
\expandafter\ifx\csname url\endcsname\relax
  \def\url#1{\texttt{#1}}\fi
\expandafter\ifx\csname urlprefix\endcsname\relax\def\urlprefix{URL }\fi
\providecommand{\bibinfo}[2]{#2}
\providecommand{\eprint}[2][]{\url{#2}}

\bibitem[{\citenamefont{Adler}(1969)}]{adler1969axial}
\bibinfo{author}{\bibfnamefont{S.~L.} \bibnamefont{Adler}},
  \bibinfo{journal}{Physical Review} \textbf{\bibinfo{volume}{177}},
  \bibinfo{pages}{2426} (\bibinfo{year}{1969}).

\bibitem[{\citenamefont{Bell and Jackiw}(1969)}]{bell1969pcac}
\bibinfo{author}{\bibfnamefont{J.~S.} \bibnamefont{Bell}} \bibnamefont{and}
  \bibinfo{author}{\bibfnamefont{R.}~\bibnamefont{Jackiw}},
  \bibinfo{journal}{Il Nuovo Cimento A (1965-1970)}
  \textbf{\bibinfo{volume}{60}}, \bibinfo{pages}{47} (\bibinfo{year}{1969}).

\bibitem[{\citenamefont{Kharzeev}(2006)}]{Kharzeev:2004ey}
\bibinfo{author}{\bibfnamefont{D.}~\bibnamefont{Kharzeev}},
  \bibinfo{journal}{Phys. Lett.} \textbf{\bibinfo{volume}{B633}},
  \bibinfo{pages}{260} (\bibinfo{year}{2006}), \eprint{hep-ph/0406125}.

\bibitem[{\citenamefont{Kharzeev et~al.}(2008)\citenamefont{Kharzeev, McLerran,
  and Warringa}}]{Kharzeev:2007jp}
\bibinfo{author}{\bibfnamefont{D.~E.} \bibnamefont{Kharzeev}},
  \bibinfo{author}{\bibfnamefont{L.~D.} \bibnamefont{McLerran}},
  \bibnamefont{and} \bibinfo{author}{\bibfnamefont{H.~J.}
  \bibnamefont{Warringa}}, \bibinfo{journal}{Nucl.Phys.}
  \textbf{\bibinfo{volume}{A803}}, \bibinfo{pages}{227} (\bibinfo{year}{2008}),
  \eprint{0711.0950}.

\bibitem[{\citenamefont{Fukushima et~al.}(2008)\citenamefont{Fukushima,
  Kharzeev, and Warringa}}]{Fukushima:2008xe}
\bibinfo{author}{\bibfnamefont{K.}~\bibnamefont{Fukushima}},
  \bibinfo{author}{\bibfnamefont{D.~E.} \bibnamefont{Kharzeev}},
  \bibnamefont{and} \bibinfo{author}{\bibfnamefont{H.~J.}
  \bibnamefont{Warringa}}, \bibinfo{journal}{Phys.Rev.}
  \textbf{\bibinfo{volume}{D78}}, \bibinfo{pages}{074033}
  (\bibinfo{year}{2008}), \eprint{0808.3382}.

\bibitem[{\citenamefont{Kharzeev et~al.}(2016)\citenamefont{Kharzeev, Liao,
  Voloshin, and Wang}}]{Kharzeev:2015znc}
\bibinfo{author}{\bibfnamefont{D.~E.} \bibnamefont{Kharzeev}},
  \bibinfo{author}{\bibfnamefont{J.}~\bibnamefont{Liao}},
  \bibinfo{author}{\bibfnamefont{S.~A.} \bibnamefont{Voloshin}},
  \bibnamefont{and} \bibinfo{author}{\bibfnamefont{G.}~\bibnamefont{Wang}},
  \bibinfo{journal}{Prog. Part. Nucl. Phys.} \textbf{\bibinfo{volume}{88}},
  \bibinfo{pages}{1} (\bibinfo{year}{2016}), \eprint{1511.04050}.

\bibitem[{\citenamefont{Kharzeev et~al.}(2013)\citenamefont{Kharzeev,
  Landsteiner, Schmitt, and Yee}}]{Kharzeev:2013jha}
\bibinfo{author}{\bibfnamefont{D.}~\bibnamefont{Kharzeev}},
  \bibinfo{author}{\bibfnamefont{K.}~\bibnamefont{Landsteiner}},
  \bibinfo{author}{\bibfnamefont{A.}~\bibnamefont{Schmitt}}, \bibnamefont{and}
  \bibinfo{author}{\bibfnamefont{H.-U.} \bibnamefont{Yee}},
  \bibinfo{journal}{Lect. Notes Phys.} \textbf{\bibinfo{volume}{871}},
  \bibinfo{pages}{pp.1} (\bibinfo{year}{2013}).

\bibitem[{\citenamefont{Kharzeev and Zhitnitsky}(2007)}]{Kharzeev:2007tn}
\bibinfo{author}{\bibfnamefont{D.}~\bibnamefont{Kharzeev}} \bibnamefont{and}
  \bibinfo{author}{\bibfnamefont{A.}~\bibnamefont{Zhitnitsky}},
  \bibinfo{journal}{Nucl. Phys.} \textbf{\bibinfo{volume}{A797}},
  \bibinfo{pages}{67} (\bibinfo{year}{2007}), \eprint{0706.1026}.

\bibitem[{\citenamefont{Erdmenger et~al.}(2009)\citenamefont{Erdmenger, Haack,
  Kaminski, and Yarom}}]{Erdmenger:2008rm}
\bibinfo{author}{\bibfnamefont{J.}~\bibnamefont{Erdmenger}},
  \bibinfo{author}{\bibfnamefont{M.}~\bibnamefont{Haack}},
  \bibinfo{author}{\bibfnamefont{M.}~\bibnamefont{Kaminski}}, \bibnamefont{and}
  \bibinfo{author}{\bibfnamefont{A.}~\bibnamefont{Yarom}},
  \bibinfo{journal}{JHEP} \textbf{\bibinfo{volume}{01}}, \bibinfo{pages}{055}
  (\bibinfo{year}{2009}), \eprint{0809.2488}.

\bibitem[{\citenamefont{Banerjee et~al.}(2011)\citenamefont{Banerjee,
  Bhattacharya, Bhattacharyya, Dutta, Loganayagam, and
  Surowka}}]{Banerjee:2008th}
\bibinfo{author}{\bibfnamefont{N.}~\bibnamefont{Banerjee}},
  \bibinfo{author}{\bibfnamefont{J.}~\bibnamefont{Bhattacharya}},
  \bibinfo{author}{\bibfnamefont{S.}~\bibnamefont{Bhattacharyya}},
  \bibinfo{author}{\bibfnamefont{S.}~\bibnamefont{Dutta}},
  \bibinfo{author}{\bibfnamefont{R.}~\bibnamefont{Loganayagam}},
  \bibnamefont{and} \bibinfo{author}{\bibfnamefont{P.}~\bibnamefont{Surowka}},
  \bibinfo{journal}{JHEP} \textbf{\bibinfo{volume}{01}}, \bibinfo{pages}{094}
  (\bibinfo{year}{2011}), \eprint{0809.2596}.

\bibitem[{\citenamefont{Torabian and Yee}(2009)}]{Torabian:2009qk}
\bibinfo{author}{\bibfnamefont{M.}~\bibnamefont{Torabian}} \bibnamefont{and}
  \bibinfo{author}{\bibfnamefont{H.-U.} \bibnamefont{Yee}},
  \bibinfo{journal}{JHEP} \textbf{\bibinfo{volume}{08}}, \bibinfo{pages}{020}
  (\bibinfo{year}{2009}), \eprint{0903.4894}.

\bibitem[{\citenamefont{Son and Surowka}(2009)}]{Son:2009tf}
\bibinfo{author}{\bibfnamefont{D.~T.} \bibnamefont{Son}} \bibnamefont{and}
  \bibinfo{author}{\bibfnamefont{P.}~\bibnamefont{Surowka}},
  \bibinfo{journal}{Phys.Rev.Lett.} \textbf{\bibinfo{volume}{103}},
  \bibinfo{pages}{191601} (\bibinfo{year}{2009}), \eprint{0906.5044}.

\bibitem[{\citenamefont{Li et~al.}(2016)\citenamefont{Li, Kharzeev, Zhang,
  Huang, Pletikosic, Fedorov, Zhong, Schneeloch, Gu, and Valla}}]{Li:2014bha}
\bibinfo{author}{\bibfnamefont{Q.}~\bibnamefont{Li}},
  \bibinfo{author}{\bibfnamefont{D.~E.} \bibnamefont{Kharzeev}},
  \bibinfo{author}{\bibfnamefont{C.}~\bibnamefont{Zhang}},
  \bibinfo{author}{\bibfnamefont{Y.}~\bibnamefont{Huang}},
  \bibinfo{author}{\bibfnamefont{I.}~\bibnamefont{Pletikosic}},
  \bibinfo{author}{\bibfnamefont{A.~V.} \bibnamefont{Fedorov}},
  \bibinfo{author}{\bibfnamefont{R.~D.} \bibnamefont{Zhong}},
  \bibinfo{author}{\bibfnamefont{J.~A.} \bibnamefont{Schneeloch}},
  \bibinfo{author}{\bibfnamefont{G.~D.} \bibnamefont{Gu}}, \bibnamefont{and}
  \bibinfo{author}{\bibfnamefont{T.}~\bibnamefont{Valla}},
  \bibinfo{journal}{Nature Phys.} \textbf{\bibinfo{volume}{12}},
  \bibinfo{pages}{550} (\bibinfo{year}{2016}), \eprint{1412.6543}.

\bibitem[{\citenamefont{Joyce and Shaposhnikov}(1997)}]{Joyce:1997uy}
\bibinfo{author}{\bibfnamefont{M.}~\bibnamefont{Joyce}} \bibnamefont{and}
  \bibinfo{author}{\bibfnamefont{M.~E.} \bibnamefont{Shaposhnikov}},
  \bibinfo{journal}{Phys. Rev. Lett.} \textbf{\bibinfo{volume}{79}},
  \bibinfo{pages}{1193} (\bibinfo{year}{1997}), \eprint{astro-ph/9703005}.

\bibitem[{\citenamefont{Akamatsu and Yamamoto}(2013)}]{Akamatsu:2013pjd}
\bibinfo{author}{\bibfnamefont{Y.}~\bibnamefont{Akamatsu}} \bibnamefont{and}
  \bibinfo{author}{\bibfnamefont{N.}~\bibnamefont{Yamamoto}},
  \bibinfo{journal}{Phys. Rev. Lett.} \textbf{\bibinfo{volume}{111}},
  \bibinfo{pages}{052002} (\bibinfo{year}{2013}), \eprint{1302.2125}.

\bibitem[{\citenamefont{Avdoshkin et~al.}(2016)\citenamefont{Avdoshkin,
  Kirilin, Sadofyev, and Zakharov}}]{Avdoshkin:2014gpa}
\bibinfo{author}{\bibfnamefont{A.}~\bibnamefont{Avdoshkin}},
  \bibinfo{author}{\bibfnamefont{V.~P.} \bibnamefont{Kirilin}},
  \bibinfo{author}{\bibfnamefont{A.~V.} \bibnamefont{Sadofyev}},
  \bibnamefont{and} \bibinfo{author}{\bibfnamefont{V.~I.}
  \bibnamefont{Zakharov}}, \bibinfo{journal}{Phys. Lett.}
  \textbf{\bibinfo{volume}{B755}}, \bibinfo{pages}{1} (\bibinfo{year}{2016}),
  \eprint{1402.3587}.

\bibitem[{\citenamefont{Tashiro et~al.}(2012)\citenamefont{Tashiro, Vachaspati,
  and Vilenkin}}]{Tashiro:2012mf}
\bibinfo{author}{\bibfnamefont{H.}~\bibnamefont{Tashiro}},
  \bibinfo{author}{\bibfnamefont{T.}~\bibnamefont{Vachaspati}},
  \bibnamefont{and} \bibinfo{author}{\bibfnamefont{A.}~\bibnamefont{Vilenkin}},
  \bibinfo{journal}{Phys. Rev.} \textbf{\bibinfo{volume}{D86}},
  \bibinfo{pages}{105033} (\bibinfo{year}{2012}), \eprint{1206.5549}.

\bibitem[{\citenamefont{Hirono et~al.}(2015)\citenamefont{Hirono, Kharzeev, and
  Yin}}]{Hirono:2015rla}
\bibinfo{author}{\bibfnamefont{Y.}~\bibnamefont{Hirono}},
  \bibinfo{author}{\bibfnamefont{D.}~\bibnamefont{Kharzeev}}, \bibnamefont{and}
  \bibinfo{author}{\bibfnamefont{Y.}~\bibnamefont{Yin}},
  \bibinfo{journal}{Phys. Rev.} \textbf{\bibinfo{volume}{D92}},
  \bibinfo{pages}{125031} (\bibinfo{year}{2015}), \eprint{1509.07790}.

\bibitem[{\citenamefont{Yamamoto}(2016)}]{Yamamoto:2016xtu}
\bibinfo{author}{\bibfnamefont{N.}~\bibnamefont{Yamamoto}},
  \bibinfo{journal}{Phys. Rev.} \textbf{\bibinfo{volume}{D93}},
  \bibinfo{pages}{125016} (\bibinfo{year}{2016}), \eprint{1603.08864}.

\bibitem[{\citenamefont{Hattori et~al.}(2017)\citenamefont{Hattori, Hirono,
  Yee, and Yin}}]{Hattori:2017usa}
\bibinfo{author}{\bibfnamefont{K.}~\bibnamefont{Hattori}},
  \bibinfo{author}{\bibfnamefont{Y.}~\bibnamefont{Hirono}},
  \bibinfo{author}{\bibfnamefont{H.-U.} \bibnamefont{Yee}}, \bibnamefont{and}
  \bibinfo{author}{\bibfnamefont{Y.}~\bibnamefont{Yin}} (\bibinfo{year}{2017}),
  \eprint{1711.08450}.

\bibitem[{\citenamefont{Xiong et~al.}(2015)\citenamefont{Xiong, Kushwaha,
  Liang, Krizan, Hirschberger, Wang, Cava, and Ong}}]{xiong2015evidence}
\bibinfo{author}{\bibfnamefont{J.}~\bibnamefont{Xiong}},
  \bibinfo{author}{\bibfnamefont{S.~K.} \bibnamefont{Kushwaha}},
  \bibinfo{author}{\bibfnamefont{T.}~\bibnamefont{Liang}},
  \bibinfo{author}{\bibfnamefont{J.~W.} \bibnamefont{Krizan}},
  \bibinfo{author}{\bibfnamefont{M.}~\bibnamefont{Hirschberger}},
  \bibinfo{author}{\bibfnamefont{W.}~\bibnamefont{Wang}},
  \bibinfo{author}{\bibfnamefont{R.}~\bibnamefont{Cava}}, \bibnamefont{and}
  \bibinfo{author}{\bibfnamefont{N.}~\bibnamefont{Ong}},
  \bibinfo{journal}{Science} \textbf{\bibinfo{volume}{350}},
  \bibinfo{pages}{413} (\bibinfo{year}{2015}).

\bibitem[{\citenamefont{Li et~al.}(2015)\citenamefont{Li, Wang, Liu, Wang,
  Liao, and Yu}}]{li2015giant}
\bibinfo{author}{\bibfnamefont{C.-Z.} \bibnamefont{Li}},
  \bibinfo{author}{\bibfnamefont{L.-X.} \bibnamefont{Wang}},
  \bibinfo{author}{\bibfnamefont{H.}~\bibnamefont{Liu}},
  \bibinfo{author}{\bibfnamefont{J.}~\bibnamefont{Wang}},
  \bibinfo{author}{\bibfnamefont{Z.-M.} \bibnamefont{Liao}}, \bibnamefont{and}
  \bibinfo{author}{\bibfnamefont{D.-P.} \bibnamefont{Yu}},
  \bibinfo{journal}{Nature communications} \textbf{\bibinfo{volume}{6}},
  \bibinfo{pages}{10137} (\bibinfo{year}{2015}).

\bibitem[{\citenamefont{Zhang et~al.}(2015)\citenamefont{Zhang, Xu, Belopolski,
  Yuan, Lin, Tong, Alidoust, Lee, Huang, Lin et~al.}}]{zhang2015observation}
\bibinfo{author}{\bibfnamefont{C.}~\bibnamefont{Zhang}},
  \bibinfo{author}{\bibfnamefont{S.-Y.} \bibnamefont{Xu}},
  \bibinfo{author}{\bibfnamefont{I.}~\bibnamefont{Belopolski}},
  \bibinfo{author}{\bibfnamefont{Z.}~\bibnamefont{Yuan}},
  \bibinfo{author}{\bibfnamefont{Z.}~\bibnamefont{Lin}},
  \bibinfo{author}{\bibfnamefont{B.}~\bibnamefont{Tong}},
  \bibinfo{author}{\bibfnamefont{N.}~\bibnamefont{Alidoust}},
  \bibinfo{author}{\bibfnamefont{C.-C.} \bibnamefont{Lee}},
  \bibinfo{author}{\bibfnamefont{S.-M.} \bibnamefont{Huang}},
  \bibinfo{author}{\bibfnamefont{H.}~\bibnamefont{Lin}}, \bibnamefont{et~al.},
  \bibinfo{journal}{arXiv preprint arXiv:1503.02630}  (\bibinfo{year}{2015}).

\bibitem[{\citenamefont{Yang et~al.}(2015{\natexlab{a}})\citenamefont{Yang,
  Liu, Wang, Zheng, and Xu}}]{yang2015chiral}
\bibinfo{author}{\bibfnamefont{X.}~\bibnamefont{Yang}},
  \bibinfo{author}{\bibfnamefont{Y.}~\bibnamefont{Liu}},
  \bibinfo{author}{\bibfnamefont{Z.}~\bibnamefont{Wang}},
  \bibinfo{author}{\bibfnamefont{Y.}~\bibnamefont{Zheng}}, \bibnamefont{and}
  \bibinfo{author}{\bibfnamefont{Z.-a.} \bibnamefont{Xu}},
  \bibinfo{journal}{arXiv preprint arXiv:1506.03190}
  (\bibinfo{year}{2015}{\natexlab{a}}).

\bibitem[{\citenamefont{Yang et~al.}(2015{\natexlab{b}})\citenamefont{Yang, Li,
  Wang, Zhen, and Xu}}]{yang2015observation}
\bibinfo{author}{\bibfnamefont{X.}~\bibnamefont{Yang}},
  \bibinfo{author}{\bibfnamefont{Y.}~\bibnamefont{Li}},
  \bibinfo{author}{\bibfnamefont{Z.}~\bibnamefont{Wang}},
  \bibinfo{author}{\bibfnamefont{Y.}~\bibnamefont{Zhen}}, \bibnamefont{and}
  \bibinfo{author}{\bibfnamefont{Z.-a.} \bibnamefont{Xu}},
  \bibinfo{journal}{arXiv preprint arXiv:1506.02283}
  (\bibinfo{year}{2015}{\natexlab{b}}).

\bibitem[{\citenamefont{Koch et~al.}(2017)\citenamefont{Koch, Schlichting,
  Skokov, Sorensen, Thomas, Voloshin, Wang, and Yee}}]{Skokov:2016yrj}
\bibinfo{author}{\bibfnamefont{V.}~\bibnamefont{Koch}},
  \bibinfo{author}{\bibfnamefont{S.}~\bibnamefont{Schlichting}},
  \bibinfo{author}{\bibfnamefont{V.}~\bibnamefont{Skokov}},
  \bibinfo{author}{\bibfnamefont{P.}~\bibnamefont{Sorensen}},
  \bibinfo{author}{\bibfnamefont{J.}~\bibnamefont{Thomas}},
  \bibinfo{author}{\bibfnamefont{S.}~\bibnamefont{Voloshin}},
  \bibinfo{author}{\bibfnamefont{G.}~\bibnamefont{Wang}}, \bibnamefont{and}
  \bibinfo{author}{\bibfnamefont{H.-U.} \bibnamefont{Yee}},
  \bibinfo{journal}{Chin. Phys.} \textbf{\bibinfo{volume}{C41}},
  \bibinfo{pages}{072001} (\bibinfo{year}{2017}), \eprint{1608.00982}.

\bibitem[{\citenamefont{Adamczyk et~al.}(2017)}]{STAR:2017ckg}
\bibinfo{author}{\bibfnamefont{L.}~\bibnamefont{Adamczyk}} \bibnamefont{et~al.}
  (\bibinfo{collaboration}{STAR}) (\bibinfo{year}{2017}), \eprint{1701.06657}.

\bibitem[{\citenamefont{Liang and Wang}(2005)}]{Liang:2004ph}
\bibinfo{author}{\bibfnamefont{Z.-T.} \bibnamefont{Liang}} \bibnamefont{and}
  \bibinfo{author}{\bibfnamefont{X.-N.} \bibnamefont{Wang}},
  \bibinfo{journal}{Phys. Rev. Lett.} \textbf{\bibinfo{volume}{94}},
  \bibinfo{pages}{102301} (\bibinfo{year}{2005}), \bibinfo{note}{[Erratum:
  Phys. Rev. Lett.96,039901(2006)]}, \eprint{nucl-th/0410079}.

\bibitem[{\citenamefont{Betz et~al.}(2007)\citenamefont{Betz, Gyulassy, and
  Torrieri}}]{Betz:2007kg}
\bibinfo{author}{\bibfnamefont{B.}~\bibnamefont{Betz}},
  \bibinfo{author}{\bibfnamefont{M.}~\bibnamefont{Gyulassy}}, \bibnamefont{and}
  \bibinfo{author}{\bibfnamefont{G.}~\bibnamefont{Torrieri}},
  \bibinfo{journal}{Phys. Rev.} \textbf{\bibinfo{volume}{C76}},
  \bibinfo{pages}{044901} (\bibinfo{year}{2007}), \eprint{0708.0035}.

\bibitem[{\citenamefont{Becattini et~al.}(2013)\citenamefont{Becattini,
  Csernai, and Wang}}]{Becattini:2013vja}
\bibinfo{author}{\bibfnamefont{F.}~\bibnamefont{Becattini}},
  \bibinfo{author}{\bibfnamefont{L.}~\bibnamefont{Csernai}}, \bibnamefont{and}
  \bibinfo{author}{\bibfnamefont{D.~J.} \bibnamefont{Wang}},
  \bibinfo{journal}{Phys. Rev.} \textbf{\bibinfo{volume}{C88}},
  \bibinfo{pages}{034905} (\bibinfo{year}{2013}), \bibinfo{note}{[Erratum:
  Phys. Rev.C93,no.6,069901(2016)]}, \eprint{1304.4427}.

\bibitem[{\citenamefont{Becattini et~al.}(2015)\citenamefont{Becattini,
  Inghirami, Rolando, Beraudo, Del~Zanna, De~Pace, Nardi, Pagliara, and
  Chandra}}]{Becattini:2015ska}
\bibinfo{author}{\bibfnamefont{F.}~\bibnamefont{Becattini}},
  \bibinfo{author}{\bibfnamefont{G.}~\bibnamefont{Inghirami}},
  \bibinfo{author}{\bibfnamefont{V.}~\bibnamefont{Rolando}},
  \bibinfo{author}{\bibfnamefont{A.}~\bibnamefont{Beraudo}},
  \bibinfo{author}{\bibfnamefont{L.}~\bibnamefont{Del~Zanna}},
  \bibinfo{author}{\bibfnamefont{A.}~\bibnamefont{De~Pace}},
  \bibinfo{author}{\bibfnamefont{M.}~\bibnamefont{Nardi}},
  \bibinfo{author}{\bibfnamefont{G.}~\bibnamefont{Pagliara}}, \bibnamefont{and}
  \bibinfo{author}{\bibfnamefont{V.}~\bibnamefont{Chandra}},
  \bibinfo{journal}{Eur. Phys. J.} \textbf{\bibinfo{volume}{C75}},
  \bibinfo{pages}{406} (\bibinfo{year}{2015}), \eprint{1501.04468}.

\bibitem[{\citenamefont{Baznat et~al.}(2016)\citenamefont{Baznat, Gudima,
  Sorin, and Teryaev}}]{Baznat:2015eca}
\bibinfo{author}{\bibfnamefont{M.~I.} \bibnamefont{Baznat}},
  \bibinfo{author}{\bibfnamefont{K.~K.} \bibnamefont{Gudima}},
  \bibinfo{author}{\bibfnamefont{A.~S.} \bibnamefont{Sorin}}, \bibnamefont{and}
  \bibinfo{author}{\bibfnamefont{O.~V.} \bibnamefont{Teryaev}},
  \bibinfo{journal}{Phys. Rev.} \textbf{\bibinfo{volume}{C93}},
  \bibinfo{pages}{031902} (\bibinfo{year}{2016}), \eprint{1507.04652}.

\bibitem[{\citenamefont{Pang et~al.}(2016)\citenamefont{Pang, Petersen, Wang,
  and Wang}}]{Pang:2016igs}
\bibinfo{author}{\bibfnamefont{L.-G.} \bibnamefont{Pang}},
  \bibinfo{author}{\bibfnamefont{H.}~\bibnamefont{Petersen}},
  \bibinfo{author}{\bibfnamefont{Q.}~\bibnamefont{Wang}}, \bibnamefont{and}
  \bibinfo{author}{\bibfnamefont{X.-N.} \bibnamefont{Wang}},
  \bibinfo{journal}{Phys. Rev. Lett.} \textbf{\bibinfo{volume}{117}},
  \bibinfo{pages}{192301} (\bibinfo{year}{2016}), \eprint{1605.04024}.

\bibitem[{\citenamefont{Deng and Huang}(2016)}]{Deng:2016gyh}
\bibinfo{author}{\bibfnamefont{W.-T.} \bibnamefont{Deng}} \bibnamefont{and}
  \bibinfo{author}{\bibfnamefont{X.-G.} \bibnamefont{Huang}},
  \bibinfo{journal}{Phys. Rev.} \textbf{\bibinfo{volume}{C93}},
  \bibinfo{pages}{064907} (\bibinfo{year}{2016}), \eprint{1603.06117}.

\bibitem[{\citenamefont{Jiang et~al.}(2016)\citenamefont{Jiang, Lin, and
  Liao}}]{Jiang:2016woz}
\bibinfo{author}{\bibfnamefont{Y.}~\bibnamefont{Jiang}},
  \bibinfo{author}{\bibfnamefont{Z.-W.} \bibnamefont{Lin}}, \bibnamefont{and}
  \bibinfo{author}{\bibfnamefont{J.}~\bibnamefont{Liao}},
  \bibinfo{journal}{Phys. Rev.} \textbf{\bibinfo{volume}{C94}},
  \bibinfo{pages}{044910} (\bibinfo{year}{2016}), \eprint{1602.06580}.

\bibitem[{\citenamefont{Becattini et~al.}(2017)\citenamefont{Becattini,
  Karpenko, Lisa, Upsal, and Voloshin}}]{Becattini:2016gvu}
\bibinfo{author}{\bibfnamefont{F.}~\bibnamefont{Becattini}},
  \bibinfo{author}{\bibfnamefont{I.}~\bibnamefont{Karpenko}},
  \bibinfo{author}{\bibfnamefont{M.}~\bibnamefont{Lisa}},
  \bibinfo{author}{\bibfnamefont{I.}~\bibnamefont{Upsal}}, \bibnamefont{and}
  \bibinfo{author}{\bibfnamefont{S.}~\bibnamefont{Voloshin}},
  \bibinfo{journal}{Phys. Rev.} \textbf{\bibinfo{volume}{C95}},
  \bibinfo{pages}{054902} (\bibinfo{year}{2017}), \eprint{1610.02506}.

\bibitem[{\citenamefont{Karpenko and Becattini}(2017)}]{Karpenko:2016jyx}
\bibinfo{author}{\bibfnamefont{I.}~\bibnamefont{Karpenko}} \bibnamefont{and}
  \bibinfo{author}{\bibfnamefont{F.}~\bibnamefont{Becattini}},
  \bibinfo{journal}{Eur. Phys. J.} \textbf{\bibinfo{volume}{C77}},
  \bibinfo{pages}{213} (\bibinfo{year}{2017}), \eprint{1610.04717}.

\bibitem[{\citenamefont{Li et~al.}(2017)\citenamefont{Li, Petersen, Pang, Wang,
  Xia, and Wang}}]{Li:2017dan}
\bibinfo{author}{\bibfnamefont{H.}~\bibnamefont{Li}},
  \bibinfo{author}{\bibfnamefont{H.}~\bibnamefont{Petersen}},
  \bibinfo{author}{\bibfnamefont{L.-G.} \bibnamefont{Pang}},
  \bibinfo{author}{\bibfnamefont{Q.}~\bibnamefont{Wang}},
  \bibinfo{author}{\bibfnamefont{X.-L.} \bibnamefont{Xia}}, \bibnamefont{and}
  \bibinfo{author}{\bibfnamefont{X.-N.} \bibnamefont{Wang}},
  \bibinfo{journal}{Nucl. Phys.} \textbf{\bibinfo{volume}{A967}},
  \bibinfo{pages}{772} (\bibinfo{year}{2017}), \eprint{1704.03569}.

\bibitem[{\citenamefont{Aristova et~al.}(2016)\citenamefont{Aristova,
  Frenklakh, Gorsky, and Kharzeev}}]{Aristova:2016wxe}
\bibinfo{author}{\bibfnamefont{A.}~\bibnamefont{Aristova}},
  \bibinfo{author}{\bibfnamefont{D.}~\bibnamefont{Frenklakh}},
  \bibinfo{author}{\bibfnamefont{A.}~\bibnamefont{Gorsky}}, \bibnamefont{and}
  \bibinfo{author}{\bibfnamefont{D.}~\bibnamefont{Kharzeev}},
  \bibinfo{journal}{JHEP} \textbf{\bibinfo{volume}{10}}, \bibinfo{pages}{029}
  (\bibinfo{year}{2016}), \eprint{1606.05882}.

\bibitem[{\citenamefont{Saffman}(1992)}]{saffman1992vortex}
\bibinfo{author}{\bibfnamefont{P.~G.} \bibnamefont{Saffman}},
  \emph{\bibinfo{title}{Vortex dynamics}} (\bibinfo{publisher}{Cambridge
  university press}, \bibinfo{year}{1992}).

\bibitem[{\citenamefont{Ricca}(1991)}]{ricca1991rediscovery}
\bibinfo{author}{\bibfnamefont{R.~L.} \bibnamefont{Ricca}},
  \bibinfo{journal}{Nature} \textbf{\bibinfo{volume}{352}},
  \bibinfo{pages}{561} (\bibinfo{year}{1991}).

\bibitem[{\citenamefont{Vilenkin and Shellard}(2000)}]{vilenkin2000cosmic}
\bibinfo{author}{\bibfnamefont{A.}~\bibnamefont{Vilenkin}} \bibnamefont{and}
  \bibinfo{author}{\bibfnamefont{E.~P.~S.} \bibnamefont{Shellard}},
  \emph{\bibinfo{title}{Cosmic strings and other topological defects}}
  (\bibinfo{publisher}{Cambridge University Press}, \bibinfo{year}{2000}).

\bibitem[{\citenamefont{Volovik}(2009)}]{volovik2009universe}
\bibinfo{author}{\bibfnamefont{G.~E.} \bibnamefont{Volovik}},
  \emph{\bibinfo{title}{The universe in a helium droplet}}, vol.
  \bibinfo{volume}{117} (\bibinfo{publisher}{Oxford University Press},
  \bibinfo{year}{2009}).

\bibitem[{\citenamefont{Eto et~al.}(2013)\citenamefont{Eto, Hirono, Nitta, and
  Yasui}}]{Eto:2013hoa}
\bibinfo{author}{\bibfnamefont{M.}~\bibnamefont{Eto}},
  \bibinfo{author}{\bibfnamefont{Y.}~\bibnamefont{Hirono}},
  \bibinfo{author}{\bibfnamefont{M.}~\bibnamefont{Nitta}}, \bibnamefont{and}
  \bibinfo{author}{\bibfnamefont{S.}~\bibnamefont{Yasui}},
  \bibinfo{journal}{PTEP} \textbf{\bibinfo{volume}{2014}},
  \bibinfo{pages}{012D01} (\bibinfo{year}{2013}), \eprint{1308.1535}.

\bibitem[{\citenamefont{Hasimoto}(1972)}]{hasimoto1972soliton}
\bibinfo{author}{\bibfnamefont{H.}~\bibnamefont{Hasimoto}},
  \bibinfo{journal}{J. Fluid Mech} \textbf{\bibinfo{volume}{51}},
  \bibinfo{pages}{477} (\bibinfo{year}{1972}).

\bibitem[{\citenamefont{Langer and Perline}(1991)}]{langer1991poisson}
\bibinfo{author}{\bibfnamefont{J.}~\bibnamefont{Langer}} \bibnamefont{and}
  \bibinfo{author}{\bibfnamefont{R.}~\bibnamefont{Perline}},
  \bibinfo{journal}{Journal of Nonlinear Science} \textbf{\bibinfo{volume}{1}},
  \bibinfo{pages}{71} (\bibinfo{year}{1991}).

\bibitem[{\citenamefont{Kharzeev}(2014)}]{Kharzeev:2013ffa}
\bibinfo{author}{\bibfnamefont{D.~E.} \bibnamefont{Kharzeev}},
  \bibinfo{journal}{Prog. Part. Nucl. Phys.} \textbf{\bibinfo{volume}{75}},
  \bibinfo{pages}{133} (\bibinfo{year}{2014}), \eprint{1312.3348}.

\bibitem[{\citenamefont{Kozhevnikov}(1999)}]{Kozhevnikov:1999ak}
\bibinfo{author}{\bibfnamefont{A.~A.} \bibnamefont{Kozhevnikov}},
  \bibinfo{journal}{Phys. Lett.} \textbf{\bibinfo{volume}{B461}},
  \bibinfo{pages}{256} (\bibinfo{year}{1999}), \eprint{hep-ph/9908444}.

\bibitem[{\citenamefont{Kozhevnikov}(2015)}]{Kozhevnikov:2015oga}
\bibinfo{author}{\bibfnamefont{A.~A.} \bibnamefont{Kozhevnikov}},
  \bibinfo{journal}{Phys. Lett.} \textbf{\bibinfo{volume}{B750}},
  \bibinfo{pages}{122} (\bibinfo{year}{2015}), \eprint{1509.01937}.

\bibitem[{\citenamefont{Holm and Stechmann}(2004)}]{holm2004hasimoto}
\bibinfo{author}{\bibfnamefont{D.~D.} \bibnamefont{Holm}} \bibnamefont{and}
  \bibinfo{author}{\bibfnamefont{S.~N.} \bibnamefont{Stechmann}},
  \bibinfo{journal}{arXiv preprint nlin/0409040}  (\bibinfo{year}{2004}).

\bibitem[{\citenamefont{Fukumoto and Miyazaki}(1991)}]{fukumoto1991three}
\bibinfo{author}{\bibfnamefont{Y.}~\bibnamefont{Fukumoto}} \bibnamefont{and}
  \bibinfo{author}{\bibfnamefont{T.}~\bibnamefont{Miyazaki}},
  \bibinfo{journal}{Journal of fluid mechanics} \textbf{\bibinfo{volume}{222}},
  \bibinfo{pages}{369} (\bibinfo{year}{1991}).

\bibitem[{\citenamefont{Kambe}(2004)}]{kambe2004geometrical}
\bibinfo{author}{\bibfnamefont{T.}~\bibnamefont{Kambe}},
  \emph{\bibinfo{title}{Geometrical theory of dynamical systems and fluid
  flows}}, \bibinfo{number}{23} (\bibinfo{publisher}{World Scientific},
  \bibinfo{year}{2004}).

\bibitem[{\citenamefont{Hirota}(1973)}]{hirota1973exact}
\bibinfo{author}{\bibfnamefont{R.}~\bibnamefont{Hirota}},
  \bibinfo{journal}{Journal of Mathematical Physics}
  \textbf{\bibinfo{volume}{14}}, \bibinfo{pages}{805} (\bibinfo{year}{1973}).

\bibitem[{\citenamefont{Ricca}(1992)}]{ricca1992physical}
\bibinfo{author}{\bibfnamefont{R.~L.} \bibnamefont{Ricca}},
  \bibinfo{journal}{Physics of Fluids A: Fluid Dynamics}
  \textbf{\bibinfo{volume}{4}}, \bibinfo{pages}{938} (\bibinfo{year}{1992}).

\bibitem[{\citenamefont{Da~Rios}(1906)}]{da1906motion}
\bibinfo{author}{\bibfnamefont{L.}~\bibnamefont{Da~Rios}},
  \bibinfo{journal}{Rend. Circ. Mat. Palermo} \textbf{\bibinfo{volume}{22}},
  \bibinfo{pages}{117} (\bibinfo{year}{1906}).

\bibitem[{\citenamefont{Kambe and Takao}(1971)}]{kambe1971motion}
\bibinfo{author}{\bibfnamefont{T.}~\bibnamefont{Kambe}} \bibnamefont{and}
  \bibinfo{author}{\bibfnamefont{T.}~\bibnamefont{Takao}},
  \bibinfo{journal}{Journal of the Physical Society of Japan}
  \textbf{\bibinfo{volume}{31}}, \bibinfo{pages}{591} (\bibinfo{year}{1971}).

\bibitem[{\citenamefont{Betchov}(1965)}]{betchov1965curvature}
\bibinfo{author}{\bibfnamefont{R.}~\bibnamefont{Betchov}},
  \bibinfo{journal}{Journal of Fluid Mechanics} \textbf{\bibinfo{volume}{22}},
  \bibinfo{pages}{471} (\bibinfo{year}{1965}).

\bibitem[{\citenamefont{Ricca}(1996)}]{ricca1996contributions}
\bibinfo{author}{\bibfnamefont{R.~L.} \bibnamefont{Ricca}},
  \bibinfo{journal}{Fluid Dynamics Research} \textbf{\bibinfo{volume}{18}},
  \bibinfo{pages}{245} (\bibinfo{year}{1996}).

\bibitem[{\citenamefont{Suzuki et~al.}(1996)\citenamefont{Suzuki, Ono, and
  Kambe}}]{PhysRevLett.77.1679}
\bibinfo{author}{\bibfnamefont{K.}~\bibnamefont{Suzuki}},
  \bibinfo{author}{\bibfnamefont{T.}~\bibnamefont{Ono}}, \bibnamefont{and}
  \bibinfo{author}{\bibfnamefont{T.}~\bibnamefont{Kambe}},
  \bibinfo{journal}{Phys. Rev. Lett.} \textbf{\bibinfo{volume}{77}},
  \bibinfo{pages}{1679} (\bibinfo{year}{1996}),
  \urlprefix\url{https://link.aps.org/doi/10.1103/PhysRevLett.77.1679}.

\bibitem[{\citenamefont{Ricca}(1994)}]{ricca1994effect}
\bibinfo{author}{\bibfnamefont{R.~L.} \bibnamefont{Ricca}},
  \bibinfo{journal}{Journal of Fluid Mechanics} \textbf{\bibinfo{volume}{273}},
  \bibinfo{pages}{241} (\bibinfo{year}{1994}).

\bibitem[{\citenamefont{Thomson}(1880)}]{thomson1880xxiv}
\bibinfo{author}{\bibfnamefont{W.}~\bibnamefont{Thomson}},
  \bibinfo{journal}{The London, Edinburgh, and Dublin Philosophical Magazine
  and Journal of Science} \textbf{\bibinfo{volume}{10}}, \bibinfo{pages}{155}
  (\bibinfo{year}{1880}).

\end{thebibliography}

\end{document}